\documentclass[a4paper,10pt]{article}

\usepackage{epsfig}
\usepackage{eurosym}
\usepackage{graphicx}
\usepackage{listings}
\usepackage{calc}
\usepackage{multirow}

\begin{document}

\title{JANUS: an FPGA-based System for High Performance Scientific Computing}

\maketitle

\begin{abstract}

This paper describes JANUS, a modular massively parallel and reconfigurable
FPGA-based computing system. Each JANUS module has a computational core and a
host. The computational core is a 4x4 array of FPGA-based processing elements
with nearest-neighbor data links. Processors are also directly connected to an
I/O node attached to the JANUS host, a conventional PC. JANUS is tailored for,
but not limited to, the requirements of a class of hard scientific applications
characterized by regular code structure, unconventional data
manipulation instructions and not too large data-base size.
We discuss the architecture of this configurable machine, and focus on
its use on Monte Carlo simulations of statistical mechanics. 
On this class of application JANUS achieves impressive performances:
in some cases one JANUS processing element outperfoms high-end PCs by
a factor $\simeq 1000$. We also discuss the role of JANUS on other
classes of scientific applications. 
\end{abstract}

\section{Overview}

Several applications in computational physics, chemistry and biology are still
far beyond the reach of state-of-the-art computers, whenever following the
evolution of even simple dynamics equations responsible for very complex
behavior requires inordinately long execution times. This is the case, for
instance, of \emph{Spin Glasses}~\cite{young} (see later, for details):
integration over configuration space of a three-dimensional lattice system of
$50^3$ sites, requiring up to $10^{13}$ Monte Carlo steps, is still an
untreatable task. One has to collect statistics for several copies of the
system, say order $10^2$, corresponding to about $10^{20}$ Monte
Carlo spin updates (\emph{spins} are the variables sitting at the lattice
sites, see later). 

Extensive use of parallelism is the main avenue to boost computer performance.
The systems we are interested in do have a very large degree of available
parallelism. \emph{Embarassingly} parallel is the fact that several copies of
the system have to be simulated independently. State-of-the-art implementations
on traditional CPUs use this approach and reach average performances (usually
measured in terms of the spin update time) of $\simeq 1$ ns/spin
when simulating $>100$ copies at the same time, exploiting the SIMD
features available in many processors. The simulation program outlined
above would take $\simeq 10^4$ years. Even if we consider that simulations have
to be performed for several values of the control parameters, it is impossible
to usefully deploy  more than several dozens of CPU's on  \emph{one} large-scale
simulation. What is needed is to exploit the intrinsic parallelism
of the simulation of \emph{each} copy of the system to shorten
processing time. 

This problem is solved by the JANUS system, described in this paper,
that exploits
parallelism up to the limits set by available
hardware resources. Each processor within JANUS is able to perform the required 
$10^{13}$  Monte Carlo steps on one copy of the system in just several months,
so the massive system of $256$ JANUS processors that we plan to deploy in spring
2008 makes the simulation program outlined above possible in about one year
time, a viable option on human timescales.

JANUS is based on FPGAs for implementation simplicity and added flexibility, so
it is an example of a reconfigurable architecture on which a high-performance
application has been succesfully configured. It also turns out that JANUS, once
its processing elements are configured for spin-glass simulations, has several
strong commonalities with \emph{many}-core architectures (see e.g.
\cite{berkeley}) that are clearly emerging as a new trend in computer
systems. So, JANUS can be seen by the physicist as a powerful simulation
engine and by the computer scientist  as a succesful example
of a \emph{massively many}-core computing structure.

This paper is structured as follows: in section~\ref{sec:ComputRequir}, we
review the computing needs of typical algorithms of statistical
mechanics simulations, deriving a set of architectural
requirements. In section~\ref{sec:ianus-hardware} we describe the structure of
the JANUS system, that we have architected trying to match in
the most flexible way the requirements of section~\ref{sec:ComputRequir}; in
section 4 we describe the programming environment available to the JANUS user,
while section \ref{sec:SG} describes the JANUS implementation of some class of
spin-models, showing aggressive exploitation of massive
on-chip parallelism, enabled by high-bandwidth on-chip memory
structures. The paper ends with a short concluding section.

\section{Computing Requirements} \label{sec:ComputRequir}
\label{sec:CompReqs}

Spin models are ubiquitous in statistical mechanics, interesting both as
description of the properties of several condensed-matter systems and as
paradigm of complexity. They are defined in terms of (generalized)-spins,
variables defined on sites $i$ of a discrete hypercubic $D$-dimensional lattice
of linear size $L$, taking only a finite (usually small) number of values. In 
the simplest case (the Ising-like models), spin variables take two values,
$s_i=\pm1$, $i=1,\dots,N=L^D$. The Monte Carlo dynamics of these systems is
governed by simple Hamiltonian functions (energy functions). In the Ising model
the energy function is

\begin{equation}\label{eq:H_ising}
E=-\sum_{\langle i,j \rangle}J_{ij} s_i s_j
\end{equation}

\begin{figure}[hbtp]
  \begin{center}
    \includegraphics[height=0.2\textheight]{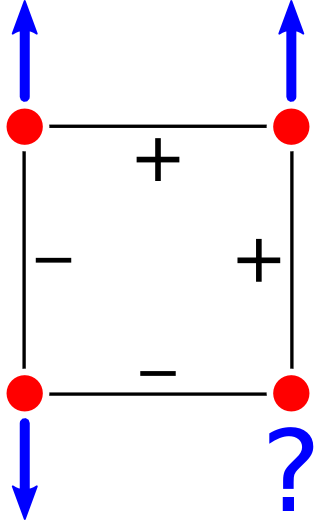}
    \hspace{3cm}
    \includegraphics[height=0.2\textheight]{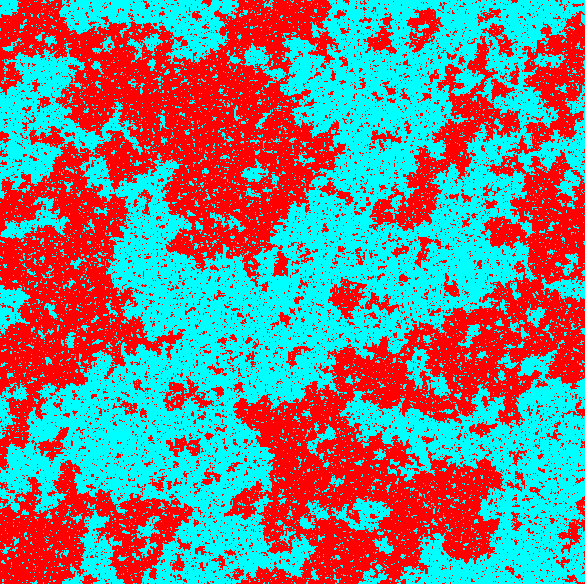}
  \end{center}
  \caption{Sketch of four neighboring nodes in a two dimensional square
    lattice (left). The two aligned spins on the topmost sites satisfy the
    positive coupling between them, as the two (misaligned) leftmost
    ones do wiht their negative coupling. The right-bottom spin cannot
    satisfy its couplings, due to a \emph{frustrated} coupling
    configuration.
    The picture on the right shows a configuration of a two
    dimensional Ising model, showing large magnetized domains. The two colors
    represent opposite spin orientations.  
  }
  \label{fig:plaq}
\end{figure}

The notation $\langle i,j \rangle$ means that summation is done only on nearest
neighbors pairs of sites on the lattice, characterized by interaction parameters
(couplings) $J_{ij}$. All symbols
in eq.~\ref{eq:H_ising} are two valued parameters. A
configuration $C_k$ is a given set of spins $s_i^{\left(k\right)}$ for all sites
on the lattice.
If all $J_{ij}$ are constant, the model describes the behaviour
of a ferromagnetic material. A set of $J_{ij}$ randomly
extracted from a bimodal distribution $J_{ij} = \pm1$ defines the
Edwards-Anderson model~\cite{young}.  Positive values of $J_{ij}$ favor aligned
spins $s_i=s_j$, while negative values favor misalignement
($s_i=-s_j$) (see Figure~\ref{fig:plaq}).

Potts models are generalization of Ising spin models: the $s_i$
are \emph{q-valued} discrete variables.
The energy function is
\begin{equation}
\label{eq:H_potts}
E=-\sum_{\langle ij \rangle} J_{ij}\delta_{s_i,s_j}
\end{equation}
where $\delta_{x,y}$ is a Kronecker delta function.
Another interesting variant is the Glassy Potts model~\cite{GPOTTS}
\begin{equation}
\label{eq:H_glassypotts}
E=-\sum_{\langle ij \rangle} \delta_{s_i,\pi_{i,j}(s_j)}
\end{equation}
where $\pi_{i,j}$ is a random permutation of $(0,1,\dots,q-1)$.

A good number of experimental systems are modeled as  Potts
systems with $q = 4$~\cite{WU}; among them
magnetic materials and absortion of molecules on metallic substrates;
the interested reader may find examples
in~\cite{ABSOR_DOMANY,ABSOR_PARK,POTTS2D_EXP,FCC_DOMANY}.

Surface growth problems in $D+1$ dimensions are also directly mappable
to chiral Potts models
\begin{equation}
\label{eq:H_chiralpotts}
E=-\sum_{\langle ij \rangle}\delta_{1+s_i,s_j}
\end{equation}
on regular D-dimensional lattices~\cite{HCS}. Surface evolution maps onto a walk
in the chiral ground states space through simple local moves, keeping the
\emph{Solid on Solid} constraint (no overhangs in the surface profile).

Potts models are relevant also in several
heterogeneous fields, even far away from magnetic and glassy materials
physics for which the above models are paradigms. Graph
coloring~\cite{GRAPH_SEM,SA} is an example 
in the broad area of optimization.
Given a graph, it consists in coloring
adjacent vertices (or, alternatively, incident edges or both incident
edges and adjacent vertices) in different colors chosen form a given
set. To achieve this goal, one can define an antiferromagnetic Potts model on top of the
graph's topology, and then the Hamiltonian 
\begin{equation} \label{eq:graph}
E=\sum_{ij \in J(G)} \delta_{s_i,s_j},
\end{equation}
where $J(G)$ is the set of the edges of the graph $G$,
which describes the number of incorrectly coloured vertices, 
since the term $\delta_{s_i, s_j}$ is $1$ if the adjacent
vertices $i$ and $j$ are of the same color. Finding a coloring which
has a null energy is therefore equivalent to a correct coloring
of the graph. Complications arising due to the
irregular structure of random graphs, will be discussed in Section~\ref{sec:SG}.

For all the models described above, knowing the energy function
determines the properties of the system.
Spin configurations are distributed according to the Boltzmann distribution:
\begin{equation}
\label{eq:boltzmann}
P(C_k) \sim e^{- \beta E(C_k)}
\end{equation}
where $\beta=1/T$ is the inverse of the temperature, describing the
probability distribution of the configurations of a system
kept at constant temperature $T$~\cite{Huang}.
Thermodynamic observables are defined as averages
weighted by the Boltzmann distribution
\begin{equation}
A=\frac{\sum_{k} A(C_k) e^{- \beta E(C_k)}}{\sum_k e^{- \beta E(C_k)}}
\label{eq:weight}
\end{equation}
where the sum runs over \emph{all} configurations. 
Full enumeration of the right hand
side of eq. \ref{eq:weight} is impossible for any reasonable lattice size.
A Monte Carlo procedure generates a sequence of configurations $C_k$
asimptotically distributed as in \ref{eq:boltzmann}, so
estimates of thermodynamic quantities become simple arithmetic averages:
\begin{equation}
A=\frac{\sum_{k} A(C_k)}{N}
\label{eq:mcmean}
\end{equation}
where $N$ is the number of configuration in the sequence. The uncertainty in
such estimate scales as $1/\sqrt{N}$.

Monte Carlo procedures come in several variants. One such variant is the
Heat Bath sequential update scheme: in the simple case of Ising models
one visits each site $i$ of the lattice and
assigns $s_i = 1$ with probability $P(s_i=+1)$ defined as follows:
\begin{eqnarray} 
P(s_i=+1) & = & f(+1)/\left[f(+1)+f(-1)\right]\ \ ,\\ 
f(s_i) & = &\exp\left[-\beta E\left(s_i\right) \right]\ \ ,\\
E(s_i) & = & -\sum_{j=\mbox{\tiny neigh}(i)}J_{ij}s_is_j - h\epsilon_is_i
\end{eqnarray}
where $E(s_i)$ is the \emph{local energy} at site $i$.

At zero temperature ($\beta \rightarrow \infty$) only energy lowering
local moves are accepted. At very high temperatures ($\beta \rightarrow 0$) we
have $P(s_i=+1) \rightarrow 1/2$ 
independent of the local energy and all spin configurations
$C_k$ are equally probable. In the intermediate (generally more
interesting) situation - finite temperature - the algorithm
samples configurations whose energies fluctuate around a value which
is a function of the temperature itself.

The Metropolis algorithm is a different Monte Carlo procedure, more efficient
than Heat-Bath for the Potts model.
The local move to update a Potts variable $s_i$ at site $i$ is made up
of three steps:
\begin{enumerate}
\item Pick up a new random value for $s^\prime_i$ 
\item compute $\Delta E=E(s^\prime_i)-E(s_i)$
\item accept the change with probability $P=\min\left\{1,\exp\left[
      -\beta \Delta E\right]\right\}$
\end{enumerate}

For any chosen Monte Carlo procedure,
the values of the couplings $J_{ij}$
determine the properties of the system. A small ferromagnetic Ising system
at zero temperature will run into one of the two
fully magnetized states (all spins aligned in the same direction). 
At finite temperature it still has two opposite
(partially) magnetized states corresponding to (free) energy minima, and
can eventually transition between tham~\cite{Huang}.
The case of the Edwards-Anderson model (randomly selected $J_{ij}=\pm
1$) is much more complex: due to concurrent couplings the energy
landscape becomes rugged, and the time needed to overcome barriers
between minima grows exponentially with system size~\cite{young}; this is the
main reason why simulations are so time consuming in these cases.

The Heat-bath algorithm for an Ising spin system is captured by the
simple (and non-optimized) code
of Listing~\ref{list:code}.
A crucial observation is that the
procedure has a very large amount of available parallelism, since $E(s_i)$
for up to one half of all spins (arranged in a checkerboard scheme)
can be computed at the same time. Up to one half of all operations
implied by Listing~\ref{list:code} can be executed in parallel, 
so all operations can 
be scheduled in two steps, if sufficient computing resources are available.
This is also true when considering other Monte Carlo update scheme,
as, for instance, in the Metropolis algorithm.

 \lstset{                  
 	basicstyle=\small,      
	showstringspaces=false, 
	language=C
}

\begin{lstlisting}[float,frame=tb,captionpos=b,label=list:code,caption={
  \small A sample non-optimized kernel of the Heat-Bath Monte Carlo simulation algorithm for the
  Edwards-Anderson model.  Variables $s$ are
  defined at all lattice sites, while $J$  couplings are defined at
  all nearest neighbor link. In this
  case the lattice side is $SIZE$ (i.e. the total number of spins is $SIZE^3$).
  HBT is a constant look-up table with 7 entries.}]
for ( k=0; k<SIZE; k++ ) {
  for ( j=0; j<SIZE; j++ ) {
    for ( i=0; i<SIZE; i++ ) {
      nbs = spin[(i+1)%SIZE][j][k] * Jx[(i+1)%SIZE][j][k] + 
            spin[(i-1)%SIZE][j][k] * Jx[(i-1)%SIZE][j][k] + 
            spin[i][(j+1)%SIZE][k] * Jy[i][(j+1)%SIZE][k] + 
            spin[i][(j-1)%SIZE][k] * Jy[i][(j-1)%SIZE][k] + 
            spin[i][j][(k+1)%SIZE] * Jz[i][j][(k+1)%SIZE] + 
            spin[i][j][(k-1)%SIZE] * Jz[i][j][(k-1)%SIZE] ;
      if ( rand() < HBT(nbs) ) { spin[i][j][k] = +1 } 
      else { spin[i][j][k] = -1 }
    }
  }
}
\end{lstlisting}

Several other features of Listing~\ref{list:code} are shared by the
applications we have in mind and are relevant in this context:

\begin{itemize}
\item The code kernel has a regular structure, associated to regular loops.
At each iteration the same set of operations is performed on data stored at
locations whose addresses can be predicted in advance.
\item Data processing is associated to simple logical operations performed on
bit-variables defining the spin variables, rather than on standard integer
or floating-point arithmetics on long data words.
\item The data base associated to the computation is of the order of just a few
bytes for each site (e.g., $\simeq 100$ KBytes for a grid of $48^3$ sites).
\end{itemize}

We are interested in a parallel processing system able to deliver very high
performance (order of 
magnitudes more than possible with current PCs). This can be done if we are able
to deploy a very large
fraction of the available parallelism available and to leverage
on the other features discussed above. The structure has to be
implemented using readily available technologies. From the architectural point
of view, our goal translates into two key features:

\begin{enumerate}
\item the processor must be configurable to perform exactly (and only)
the subset of functions required by the algorithm (and poorly supported
by traditional architecture) so, for a given
silicon budget, the largest possible number of operations, exploiting
the parallelism offered by the algorithm, can be performed.
\item the performance made in principle possible by the previous point must be
sustained by sufficient memory bandwidth (as we will se later,
in our target application we move from/to memory $\simeq 4000$ bits per
clock cycle). This dictates the use of fine grained on-chip memory structures.
\end{enumerate}

We make the choice of using latest generation FPGAs as the basic block
of our system. Their reconfigurable structure matches point 1 above, and the
availability of a large number of on-board RAM blocks matches point 2.
The system that we have defined and built is described next.

\section{The JANUS architecture} \label{sec:ianus-hardware}

JANUS builds on previous attempts made several years
ago within the SUE project \cite{sue}, with a much wider 
application focus and much higher expected performances, made possible by a more
flexible design and by recent progress in FPGA technology (for early ideas on
JANUS, see \cite{janus}).

\begin{figure}[hbtp]
  \begin{center}
    \includegraphics[width=0.35\textwidth]{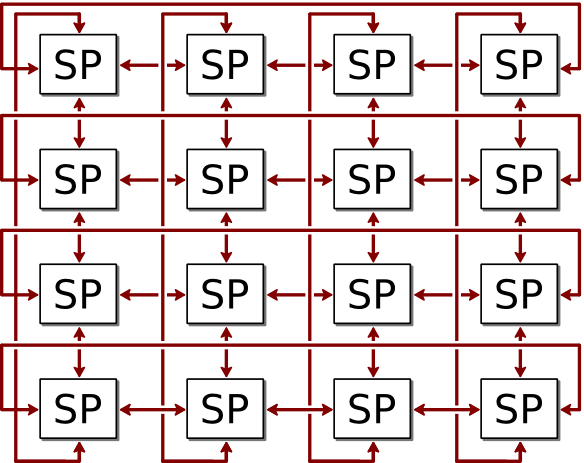}
    \includegraphics[width=0.45\textwidth]{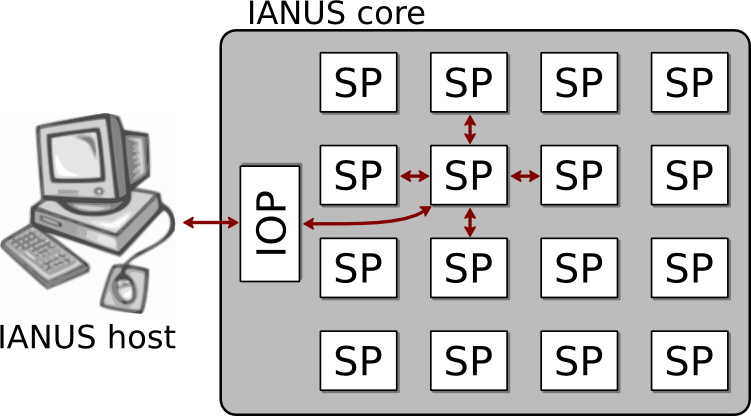}
  \end{center}
  \caption{\small JANUS core: topology (left) and overview of the JANUS
      prototype implementation (right) with 16 SPs, the I/O module (IOP) and the JANUS host.}
  \label{fig:giorgio}
\end{figure}

\subsection{Overall structure}

JANUS is based on modules containing a computational core (the JANUS core)
loosely connected to a host system (the JANUS host) of one or more networked
PCs (see Figure~\ref{fig:giorgio}).

The JANUS core is based on:
\begin{itemize}
  \item a $4 \times 4$ 2-D grid of processing elements
    (each called SP) with nearest-neighbor links in both axis 
    (and periodic boundary conditions), making up the number-crunching kernel of
    the system;
  \item an IO-Processor (IOP) connected with all SPs and to the host.
\end{itemize}
Both the SPs and the IOP are Xilinx Virtex4-LX200 FPGAs.

Each SP processor contains uncommitted logic that can be configured at will by
the IOP. Once configured, the SP is the main computational engine of the system.
Number crunching kernels are allocated to one or more SPs. They run either as
independent threads, ultimately reporting to the JANUS host through the IOP or
they run concurrently, sharing data, as appropriate, across the nearest-neighbor
links (in the current implementation each full-duplex link transfers four bytes
per clock cycle with negligible latency, i.e. one or two clock cycles). A single
task may execute on the complete core, or up to 16 independent tasks may be
allocated on the system.
A major design decision has been that the SPs have no memory apart from the one
available on-chip (of the order of $\simeq 0.5$ MByte per processing element).
This choice constrains the class of problem that JANUS can handle. On the other
hand, in their quest for performance, applications can rely on the huge
bandwidth and low latency made possible by on-chip memory to feed all the cores
configured onto the FPGA.

The IOP module controls the operation of all available SPs, merges and moves
data to/from the JANUS host on full-duplex IOP-SP point-to-point link
transfering 2 bytes of data every clock cycle. A bank of staging memory  is
available, and the main communication channel uses 2 Gigabit-Ethernet  links (an
USB port and a serial interface are available for debugging and test).

The IOP FPGA is configured  as a relatively stable control and I/O interface for
the SPs. However close to $70\%$ of the available logical resources are
left over, so more complex roles within large, multi-SP applications
are possible. The IOP might act as a cross-bar switch for the array
of SPs, supporting complex patterns of data transfers or perform global
operations on aggregated data sets.  We also plan  to embed a standard
microprocessor (e.g. MicroBlaze\texttrademark) in order to support more complex data
handling. In principle a slim Linux-like operating system might be supported,
making the JANUS unit an independent computer.

\begin{figure}[hbtp] 
  \begin{center}
    \includegraphics[width=(0.5\textwidth)]{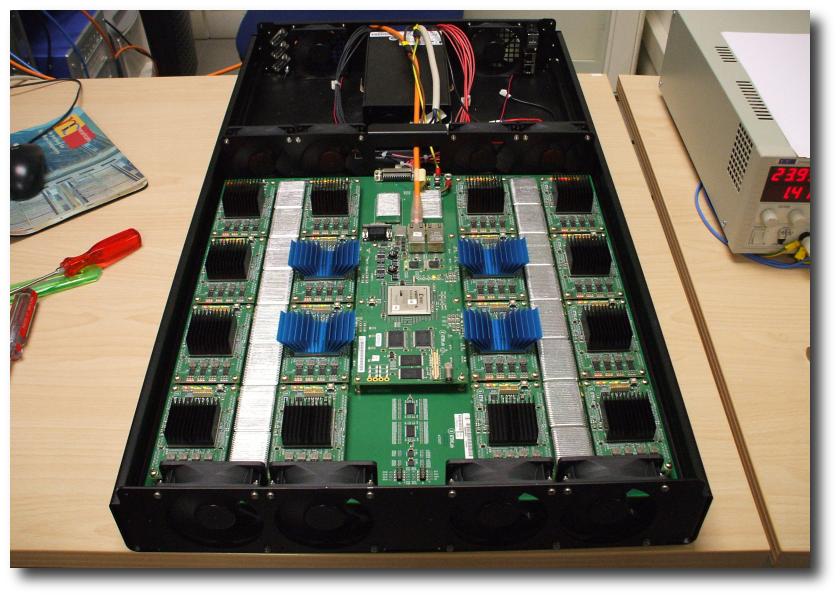}
    \includegraphics[width=(0.3\textwidth)]{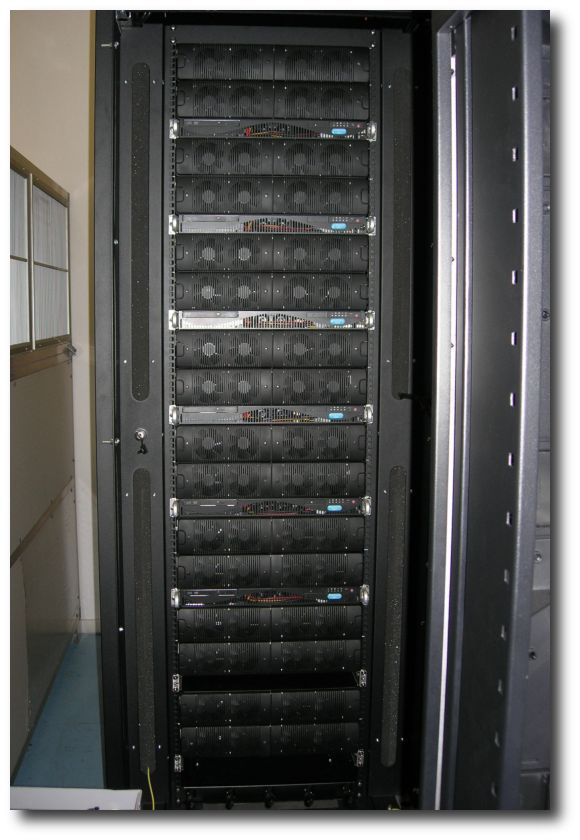}
  \end{center}
    \caption{\small A JANUS module (left) and the JANUS rack (right) with
      16 JANUS modules and 8 JANUS hosts.}
  \label{fig:fotoDaniele}
\end{figure}

\subsection{IOP firmware}
\label{sec:iopStruct} 

The current firmware configuration of the Input/Output Processor (IOP) focuses
on the implementation of an interface between the SPs and the host. The IOP is
not a programmable processor with a standard architecture: it basically allows 
streaming of data from the host
to the appropriate destination (and back), under the complete control of the
JANUS host. 

The IOP configured structure is naturally split in 2 segments:
\begin{itemize}

\item the \emph{IOlink} block handles the I/O interfaces between IOP
  and the JANUS host (gigabit channels, serial and USB ports). Its
  supports the lower layers of the gigabit  ethernet protocol, ensures
  data integrity and provides a bandwidth close to the theoretical
  limit. 

\item the \emph{multidev} block contains a logical device associated to
each hardware sub-system that may be reached for control and/or data transfer:
there is a memory interface for the staging memory, a programming interface to
configure the SPs, an SP interface to handle
communication with the SPs (after they are configured) and several service/debug
interfaces.
\end{itemize}

The link between these two segments is the \emph{Stream Router} that
forwards the streams coming from the IOlink to the appropriate device within
the \emph{multidev} according to a mask encoded in the stream.

\begin{figure}[hbtp]
  \centering
  \includegraphics[width=0.98\textwidth]{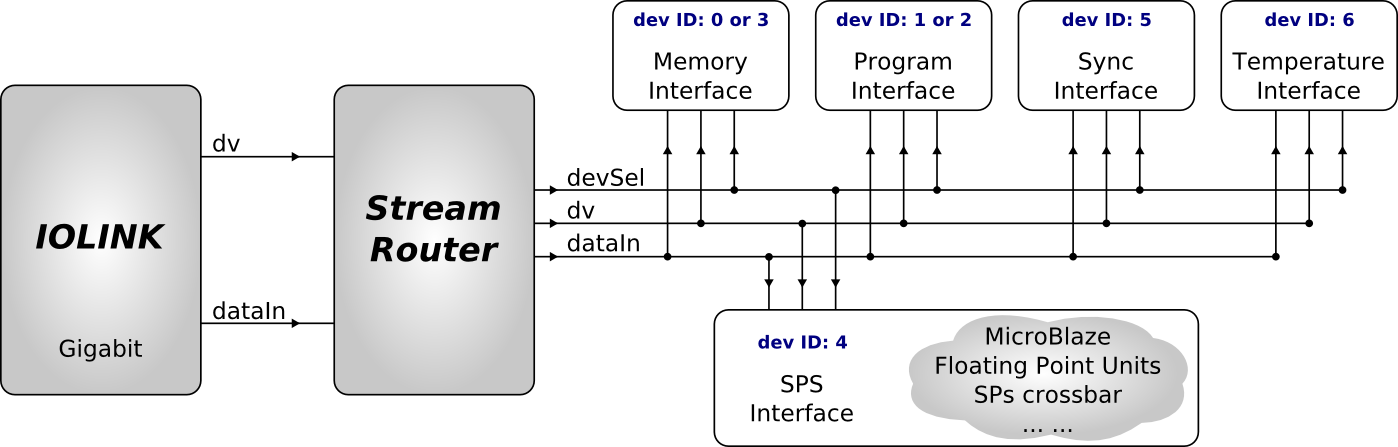}
  \caption{\small Schematic logical representation of the IOP processor: 
  on the left the \emph{IOlink} that handles the I/O interfaces between IOP and 
  the JANUS host; on the right the devices of the \emph{multidev}.}
  \label{fig:IOP}
\end{figure}

A block diagram of the IOP is shown in Figure~\ref{fig:IOP}.

As shown in Figure~\ref{fig:IOP}, the SP interface plays a special role 
in that it can be configured to perform computing functions common to all the
SPs involved in a given computational thread.

The IOP supports a low-level interface mechanism between the host computer and
any configured JANUS device flexible enough to adapt to a configurable computer:
basically the IOP handles so-called \emph{data-worms}: each worm is just a sequence
of commands and data that is not interpreted by the IOP but simply delivered
to the configured device.

\subsection{System-level structure} 

The IOP and SP modules are engineered as small daughter-boards plugged onto a 
Processing Board (PB). This implementation will allow to upgrade each
component independently. We are already considering mid-term upgrades such as
newer generation FPGAs for the SPs and more performing core-host interfaces
(using e.g. PCI-Express).

Several JANUS modules can be grouped into a larger system: each core is packed
in one standard rack-mountable 19-inch box.  Several cores and their
associated hosts are placed in a standard rack. All JANUS hosts are networked
and connected to a supervisor host. End-users log onto the supervisor and
connect to one (or more) available JANUS modules.

After a test period using prototype boards with Virtex4 LX160 FPGAs, we are at 
the moment testing the first complete JANUS system: 16 JANUS cores with  8 JANUS
hosts assembled in a standard 19'' rack. 
Three more JANUS cores with two JANUS hosts are also available for less
compute-intensive applications and for program development and test.
Pictures of a JANUS module and of the large JANUS rack are shown in Figure
\ref{fig:fotoDaniele}.

We plan to start physics simulations on these systems in spring 2008.

\section{Software Environment}

Development of JANUS applications involves firmware design to configure the SPs,
and the related host-based software development. Firmware development is
performed by means of standard commercial design tools, starting from HDL
codes.  Handcrafting and optimizing a complex application for JANUS is
admittedly a complex and time consuming investment, justified in the statistical
physics community by huge rewards in term of performance. This trade-off is not
necessarily applicable in general; at this stage, our project does not plans
specific steps in the field of making reconfigurable computing more user
friendly to the general users. We plan however to make available all the
applications that will be developed by the JANUS community in the form of a
collection of tested firmware modules.

We have developed a host-resident software environment able to cope with any
SP-based firmware, as long as the latter adheres to a model in which JANUS is a
memory based coprocessing engine of the host computer; user programs move their
data onto the JANUS storage devices (FPGA embedded memories), activates JANUS
processing and, at the end of the computation, retrieve processed data.

This model is supported by an host-resident execution and development
environment that we call \emph{JOS} (JANUS Operating System). \emph{JOS} runs on
any Linux based PC and builds on a low-level C library, based on standard Unix
raw network sockets, implementing the protocol needed to communicate  with the
IOP firmware on the Gbit ethernet link.

For the application user, \emph{JOS} consists of:
\begin{itemize}

\item A set of libraries with primitives to control IOP devices, written both in
  PERL and C (\texttt{JOSlib})
\item A multi-user environment for JANUS resource abstraction and concurrent
  jobs management (\texttt{josd})
\item A set of SP firmwares for several scientific applications and the C
  libraries needed to control them via the \texttt{josd} environment
  (\texttt{jlib}).
\end{itemize}

\texttt{josd} is a background job running on the JANUS host, providing
hardware abstraction and an easy interface to user applications. It
hides all details of the underlying structure of JANUS, mapping it to
the user as a simple grid of SPs. It interfaces to the user only via
the Unix socket API, so high-level applications may be written in
virtually any language.

For debugging and test, an interactive shell (\texttt{JOSH}), written in PERL
and also based on (\texttt{JOSlib}) offers expert users complete access to
all JANUS resources. The JOSH interactive shell provides direct
access to JANUS resorces, allowing to communicate with the IOP and drive all its
internal devices. Whenever a new SP firmware is developed, new primitives to
instruct the SP interface of the IOP are added in \texttt{JOSlib}.

\section{Reconfiguring JANUS for spin-system simulations}  \label{sec:SG}
SP configurations for the simulation of three dimensional Ising spin models,
defined by a more general version of Eq.~\ref{eq:H_ising} (including the
site-diluted case), as well as for the standard and glassy Potts model
Eq.~\ref{eq:H_glassypotts} have been already developed and tested and
are discussed extensively in~\cite{janusCPC}.
We present here just a few highlights.

Consider the Metropolis Monte Carlo scheme discussed above.
Steps 1 and 3, that consume random numbers, involve
some 32-bit integer arithmetics (we choose the Parisi-Rapuano shift
register generator~\cite{rapuano}).
Step 2 involves logical (model specific) bitwise operations for each
site.
 
We divide the lattice in a checkerboard scheme, updating
all black (white) spins simultaneously and independently.
In addition, when simulating disordered
systems one usually has to deal with two replicas (independent
systems with same coupling configurations), so we can mix all
black (white) sites of one lattice system with the (white) black ones
of its replica: spins on sites of one of the two resulting
\emph{artificial} mixed replicas can be updated all
simultaneously, with all their neighbors residing in the other one.

Computations involved in step 2 can be programmed very
efficiently on the basic logical operators of the FPGA. Several
\emph{updating cells} may be instantiated to operate in parallel,
provided that the needed bandwidth from and to spin and interaction
configurations storage is guaranteed.
FPGAs comes with many embedded configurable memories (known as
\emph{Block RAMs}). Each one has a natural 2D width
$\times$ depth structure and ``stacking'' some of them we arrange
a 3D matrix of one bit variables. Taking $z$ as the coordinate ``inside''
the memories, $y$ as the index of ``stacked'' memories and
$x$ along their width, an entire $(xy)$ plane may be fetched and fed to
update logic by simply addressing all memories with the same $z$.
The total number of these memory structures depends on the model: we
instantiate two such structures (we have two checkerboard mixed
replicas) for each bit needed to represent spin 
variables, and one structure for each bit needed to represent
couplings ($1$ for the Edwards-Anderson and the disordered Potts
models, $8$ for random permutations in $4$-states Glassy Potts model,
and three directions in 3D lattices must be considered).
Reminding the limit of two I/O operations on each Block RAM per clock
cycle, processing rate in which an entire plane of one of the two mixed
replicas  can be updated at each clock cycle can be easily sustained,
as to process the $z$ plane we only need the same $z$ plane of the
interaction memory structure, together with planes $z-1$, $z$ and
$z+1$ of the neighbors mixed lattice. So, we end up with a pipeline in which we
fetch $z+1$ from memory at the same time as we write $z-1$.

A number of update cells equal to the number of sites in a plane are
implemented in configurable logic. Each cell receives an updating spin,
its neighbors and a new random spin value as input, 
along with all needed couplings, and
computes an integer representations of the energy change. The latter becomes
the pointer to a small (tipically not more than 13 values) 
look-up table containing transition
probabilities, also implemented in configurable logic as distributed
RAM. Considering the limit of two I/O per clock, each look-up table
can be shared by two update cells. Transition probabilities are
represented as integers, so they can be immediately
compared to random numbers to test for acceptance in the Metropolis
procedure.
As discussed in~\cite{janusCPC}, we implement the Parisi-Rapuano shift
register random number generator in a very effective way, allowing for
the generation of hundreds of $32$-bit random numbers at each clock
cycle. 

Available on-chip memory limits Ising models to lattice sizes not
larger than $96^3$ and 4-states disordered Potts models to
$88^3$. Available configurable logic limits the number of parallel
updates to $1024$ for Ising spin models and $512$ for 4-states Potts models. 

Table~\ref{tab:perfs} summarize performances on one SP (in terms of the
update time for a single spin)  and compares with high-end PCs. We comment on
these figures below.

Things are slightly more complicated for graph coloring.
Remember that no adjacent vertices can be updated in parallel,
using the Metropolis algorithm. One way to avoid this
problem is to perform a preliminary reordering of the graph vertices in
$P$ subsets of non adjacent vertices, and then updating the elements
of each $P$ in parallel. It is possibile to do this in various ways,
topological or not. This reordering is in principle also a hard
problem. It is however not a bottleneck in our case as we mainly
consider low-connectivity graphs, and we are only interested in
reasonable (not necessarily optimal) solutions.

We perform graph reordering on a standard pc and then run an optimized parallel
update routine on JANUS.
A challenging memory handling problem arises, as an irregular graph structure translates
into irregular memory access patterns.

\begin{figure}[h]
  \begin{center}
    \includegraphics[width=0.8\textwidth]{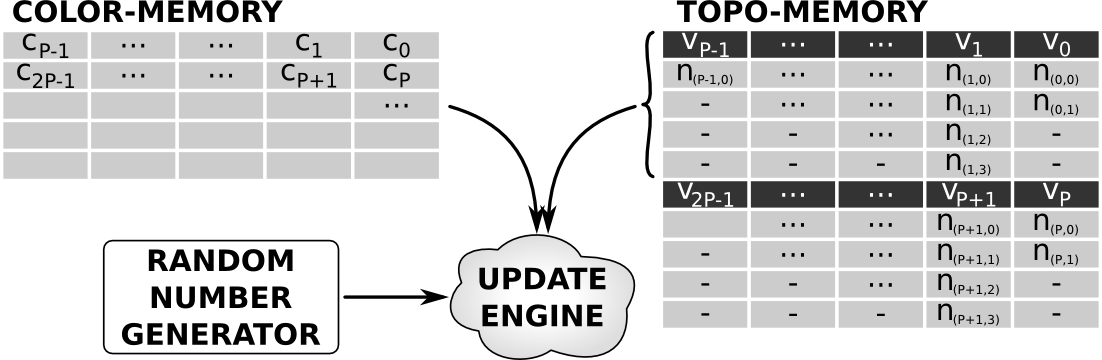} 
    \caption{\small The graph coloring scheme on the JANUS' SP: the
      graph topology is stored in the TOPO-memory (TM) while the color
      of each vertex is stored in another memory, called COLOR-memory
      (CM).}
  \end{center}
  \label{fig:graph}
\end{figure}

We assume that each vertex $V$ has an index/address $V_i$ (i.e. a
``label'' to localize it in the graph) and a color $C_i$. We
map the graph on two different structures: a topological
structure (TM) contanining pointers to the
neighbors of each vertex and a second structure (CM) storing current values 
for the vertex colors. 

These structures are sketched in Figure~\ref{fig:graph}, where
each row represents one memory word. Inside CM a
row stores the colors of P non-adjacent vertices. In the TM we store
in the black rows the pointer to each P vertex and in the gray
columns the pointer of each vertex that is neighbor of the
vertex in the black row up. CM is replicated $P/2$ times so all needed accesses
can be performed at each clock cycle. 

In our preliminary tests we try to color with $Q = 3$ or $4$ colors
graphs of $\approx16000$ vertices with average connectivity $C_m =
4$. A high-end PC (Intel Core 2 Duo, 64
bit, 4 MB cache) updates one vertex in $27 ns$, while JANUS does the same in
$\approx 2.7 ns$. 
Performances in this case are good but not so impressive as in the previous
models: irregular memory access and small on-chip memory are the limiting
factors. 

\begin{table}[hbtp]
\begin{center}
\begin{tabular}{||l|c|c|c|c|c|c||}
\hline
\hline
\multicolumn{2}{|c|}{} & \multicolumn{2}{|c|}{JANUS} & \multicolumn{3}{|c|}{PC} \\
\hline
MODEL 
& Algorithm & Max size & perfs & AMSC & SMSC
& NO MSC\\
\hline
3D Ising EA &  
Metropolis & $96^3$ & 16 ps & $45\times$ & $190\times$  &  \\
3D Ising EA &  
Heat Bath  & $96^3$ & 16 ps & $60\times$ &         &  \\
$Q=4$ 3D Glassy Potts     & 
Metropolis & $16^3$ & 64 ps & $1250\times$  & $1900\times$  &  \\
$Q=4$ 3D disordered Potts &
Metropolis & $88^3$ & 32 ps & $125\times$   &         & $1800\times$\\
$Q=4$, $C_m=4$ random graph &
Metropolis & 24000 & 2.5 ns & $2.4\times$   &         & $10\times$  \\
\hline
\hline
\end{tabular}
\end{center}
\caption{Summary of performances on different models, in terms of the
  time needed to update a single spin, on one JANUS SP and an Intel
  Core 2 Duo 2.0 GHz. Times for PC are given as multiples of the
  corresponding FPGA
  update times. Maximum lattice sizes are reported in the FPGA
  case. PC performances do not depend on lattice sizes as long as
  all data structures can be kept only in cache. Performances for each
  JANUS core are 16 times better.}
\label{tab:perfs}
\end{table}

We now come to a comparison with conventional systems.
Traditional architectures have two bottlenecks: it is impossible to generate
many random numbers at the same time while logical manipulations of independent
bit-valued variables waste hardware resources. Hence, the common trick to resort
to multi-spin coding (MSC) in which several (short) variables are mapped on one
(long) machine word. In an asynchronous scheme (AMSC) the variables mapped on
each word are associated to \emph{different} copies of the system. In this case
the same random number can be used to control all updates performed in parallel,
boosting performance. On the other hand, as discussed earlier, we need to
simulate efficiently \emph{one} (or a relatively small number of) system. In
order to do so, synchronous schemes (SMSC) code in each machine word more sites
of the  \emph{same} system. In this case the bottleneck is the generation of
many random numbers (one for each mapped site; in the Metropolis case, the
management of the probability tables also becomes much more complex):
performance is much lower.

Performance comparison for all models is done in Table~\ref{tab:perfs};
In most cases performance is outstanding.
It is striking to note the difference in performance for traditional
architectures between Ising and Potts models, i.e., between two different but
vastly similar systems. This is due to the fact that one resorts to tricks to
perform operations that are not supported by typical native instruction sets: in
some cases effective tricks can be found in slightly different situation the
same trick may not work.

This is the area where reconfigurability offers a dramatic performance
boost, as we setup data-paths that use all available resources to
perform exactly the required mix of operations required by the algorithm.
Our most effective handles are the limited complexity of the logic manipulation
of the spin-variables and the possibility to instance massively parallel random
number generators. Resources are assigned to the two main task in a well
balanced way. 
The additional complexity of Potts models respect to the Ising spin ones
results in increasing connectivity between FPGA internal devices, but
it is completely taken up in increasing design flow run times (especially
placing and routing), that may be considered equivalent to
\emph{compilation  time} on standard computers.

The more impressive boost obtained in FPGA respect to PCs is in
the Glassy Potts case, as the need of many bits to represent random
permutations invalidates most of the benefits of multi-spin coding
techniques.

Our level of performance is supported by the required memory
access bandwidth, that we obtain by a judicious data layout,
supported in hardware by the fine-grained memory structure within the FPGA. 

Irregular access patterns is the main reason that
makes the comparison between JANUS and PCs less striking in the graph
coloring problem. We try to escape this problem using multiple copies of the
system, but this is resource greedy, so we are hit by limited on-board memory
size. In general this is a reminder that memory bottlenecks are more and more
a performance problem. 

  \section{Conclusions}
In this paper we have described the JANUS computer, a parallel system
that explores the potential for performance of FPGAs on massively
parallel algorithms. 

We have tested JANUS performance 
on a class of applications relevant for statistical
physics. Our results rest on instantiating available
parallelism in an efficient way on the logical resources of
the FPGA. Fine-grained memory structures in turn
provide the huge memory bandwidth required to sustain performance.

We considered a large but limited class of
applications for which a careful handcrafting of the kernel codes is required.
Some more general lessons can be drawn:

\begin{itemize} 
\item FPGAs can be configured to perform functions poorly done by
traditional CPUs, with exceptional performance rewards.
\item The main performance bottleneck is the amount of available on-chip
  memory: as long as the simulation database can be stored on
  internal RAM blocks, enough bandwidth can be provided to support
  performance. On the other hand, as soon as the database becomes so large that
  (lower-bandwidth) external memory is needed, performance drops significantly.
\item The problem is made more serious when access patterns are irregular (for
graph coloring in our case, but a common problem in many other cases). FPGA
families
oriented to high-performance computing, with larger memories (at the expense
of configurable logic) might be a partial solution to the problem.
\end{itemize}

It would be interesting to compare the performances of JANUS with those
obtainable on commercial multi-core chips,
such as, for instance, the IBM \emph{Cell BE}. In principle one could
estimate that each SPE core has roughly the same performances as a
traditional processors (clock frequencies are approximately
the same and the level
of parallelism associated to SIMD instructions is the same). More
accurate evaluations require considerable additional studies, as in
many cases it has been shown that actual performances depend
critically on the size of the system to be mapped on each SPE and on
the detailed pattern of communication among SPEs, between SPEs and their
local storage  and with
external memory.~\cite{LAT2007} Work on this topic is in progress. 

\section*{Acknowledgments}
We thank M. Lena and S. Sialino for their outstanding 
support during the commissioning of the first JANUS system.

\end{document}